\documentclass[aps,pra,showpacs,twocolumn,superscriptaddress, floatfix, nofootinbib]{revtex4-2}%

\usepackage[utf8]{inputenc}
\usepackage{epsfig}
\usepackage{url}
\usepackage{multirow}
\usepackage{makecell,booktabs} 
\usepackage{capt-of} 

\usepackage{graphicx}
\usepackage{tikz} 

\usepackage{amsfonts}
\usepackage{amssymb}
\usepackage{amsmath}
\usepackage{verbatim}
\usepackage{listings}
\usepackage{float} 
\usepackage{bm} 
\usepackage[qm,braket]{qcircuit} 

\usepackage{mathtools}

\usepackage{color,xcolor}

\usepackage{hyperref}
\hypersetup{  colorlinks=true, linkcolor=blue, citecolor=red, urlcolor=blue  }

\usepackage{physics}

\renewcommand{\bf}[1]{\mathbf{#1}}

\newcommand{\bb}[1]{\mathbb{#1}}

\renewcommand{\eqref}[1]{Eq.~(\ref{#1})}

\usepackage{amsthm}
\newtheorem{theorem}{Theorem}

\theoremstyle{definition}

\usepackage{txfonts}

\DeclareMathOperator{\poly}{poly}

\newcommand{\p}{\vb{p}}
\renewcommand{\P}{\mathcal{P}}
\newcommand{\s}{\vb{s}}
\newcommand{\x}{\vb{x}}
\renewcommand{\a}{\vb{a}}
\newcommand{\y}{\vb{y}}

\renewcommand{\c}{\vb{c}}
\newcommand{\Z}{\mathcal{Z}}
\newcommand{\C}{\mathcal{C}}

\DeclareMathOperator{\row}{row}
\DeclareMathOperator{\col}{col}

\begin{document}

\title{Anti-Forging Quantum Data: Cryptographic Verification of Quantum Computational Power}

\author{Man-Hong Yung}
\email{yung@sustech.edu.cn}
\affiliation{Department of Physics, Southern University of Science and Technology, Shenzhen 518055, China.}
\affiliation{Shenzhen Institute for Quantum Science and Engineering, Southern University of Science and Technology, Shenzhen 518055, China.}
\affiliation{Guangdong Provincial Key Laboratory of Quantum Science and Engineering, Southern University of Science and Technology, Shenzhen 518055, China.}
\affiliation{Shenzhen Key Laboratory of Quantum Science and Engineering, Southern University of Science and Technology, Shenzhen, 518055, China.}

\author{Bin Cheng}
\email{chengb@mail.sustech.edu.cn}
\affiliation{Department of Physics, Southern University of Science and Technology, Shenzhen 518055, China.}
\affiliation{Shenzhen Institute for Quantum Science and Engineering, Southern University of Science and Technology, Shenzhen 518055, China.}
\affiliation{Centre for Quantum Software and Information, Faculty of Engineering and Information Technology, University of Technology Sydney, NSW 2007, Australia.}







\begin{abstract}
Quantum cloud computing is emerging as a popular model for users to experience the power of quantum computing through the internet, enabling quantum computing as a service. The question is, when the scale of the computational problems becomes out of reach of classical computers, how can users be sure that the output strings sent by the server are really from a quantum hardware? 
In 2008, Shepherd and Bremner proposed a cryptographic verification protocol based on a simplified circuit model called IQP (instantaneous quantum polynomial-time), which can potentially be applied to most existing quantum cloud platforms. However, the Shepherd-Bremner protocol has recently been shown to be insecure by Kahanamoku-Meyer. 
Here we present an extended model of IQP-based cryptographic verification protocol, where the Shepherd-Bremner construction can be regarded as a special case. This protocol not only can avoid the attack by Kahanamoku-Meyer but also provide several additional security measures for anti-forging quantum data. In particular, our protocol admits a simultaneous encoding of multiple secret strings, strengthening significantly the hardness for classical hacking. Furthermore, we provide methods for estimating the correlation functions associated with the secret strings, which are the key elements in our verification protocol. 

\end{abstract}

\maketitle



\section{Introduction}
Despite the fact that near-term quantum computers would be noisy and of intermediate scale~\cite{Preskill2018}, they have the potential in performing specific tasks intractable for any classical computer, a status knows as quantum supremacy~\cite{Preskill2012,Lund2017,Terhal2018,YungNSR,arute_quantum_2019}. A natural question arises: how could we tell if a remote device is truly quantum or not? This question is not only of fundamental interest in nature~\cite{aharonov_is_2012}, but also relevant to many computing protocols involving two parties, namely a verifier and a prover.

This question is called verification of quantum computational power, or test of quantumness, and it is a simplified version of the verification of \emph{any} quantum computation.
In the setting of the general problem, a quantum prover interacts with a classical verifier, in order to convince the verifier that the results are correct for the given problem in \textsf{BQP};
here, \textsf{BQP} is the class of decision problems that a quantum computer can efficiently solve.
Verification protocols for the general problem can be achieved if certain relaxation is allowed.
For example, one may assume that the verifier can use limited quantum computational power and quantum communication~\cite{Broadbent2008,Broadbent2010-blind-original,Aharonov2008,Aharonov2017, Fitzsimons2017-verifiable-blind, Fitzsimons2018, Mills2018-it-secure}, or one may allow multiple spatially separated entangled provers ~\cite{Reichardt2013-blind-vazirani, Fitzsimons2018}. These protocols can be made secure without any computational assumption. 
However, the caveat is that they may not be applicable to the near-term quantum cloud computing model in practice. 

On the other hand, with cryptographic assumptions of trapdoor claw-free functions (TCF), such as the quantum hardness of the learning-with-errors (LWE) problem~\cite{regev2009lwe}, Mahadev devised a four-message verification protocol, involving a purely classical verifier and a single untrusted quantum prover~\cite{Mahadev2018}.
Mahadev's protocol was later improved to a two-message protocol via the Fiat-Shamir transform~\cite{alagic_two-message_2019}. 
Back to the problem of verifying quantum computational power, several verification protocols were proposed based on similar cryptographic assumptions of TCFs as in Mahadev's protocol~\cite{brakerski_cryptographic_2018, brakerski_simpler_2020, kahanamoku-meyer_classically-verifiable_2021}.
However, implementing these TCF-based protocols requires thousands of qubits to ensure security, far out of reach of current quantum technology.

Specifically, existing quantum cloud computing models involve purely classical clients who can only send out classical descriptions of the quantum circuits to the service provider and receive the resulting statistics of the output bit strings through the internet. 
Again, from the practical point of view, it would be desirable to directly incorporate the verification process into the cloud computing process without much modifications.
In 2008, based on the model of IQP (instantaneous quantum polynomial-time) circuits, Shepherd and Bremner proposed~\cite{IQP08} a cryptographic verification protocol which fits very well the scope of the near-term quantum cloud computing. 
In their protocol, the verifier constructs an IQP circuit from the quadratic-residue code (QRC)~\cite{macwilliams1977book}, which is encoded with a secret string kept by the verifier. 
Then, he/she sends the classical description of the IQP circuit to the prover, who returns the output strings of the IQP circuit to the verifier. Finally, the verifiers checks if a certain measurement probability, associated with the secret string, is the same as a pre-determined value. 

The Shepherd-Bremner protocol was previously believed to be secure for a long time, partly because of the classical intractability of simulating general IQP circuits~\cite{IQP10, bremner_average-case_2016}; generally, IQP circuits cannot be sampled even approximately with efficient classical algorithms, unless the polynomial hierarchy collapses, which is a highly implausible complexity-theoretic consequence.
However, Kahanamoku-Meyer has recently found~\cite{kahanamoku-meyer_forging_2019} a loophole in the Shepherd-Bremner protocol, and devised a classical algorithm to break the Shepherd-Bremner protocol. The loophole is originated from the special properties of QRC encoding the secret string from the IQP circuit. Once the secret is known, the prover can efficiently generate output strings that can pass the verification test, even without running the quantum circuit. 

Here, we present a generalized model of IQP-based cryptographic verification protocol for near-term quantum cloud computing, which can be reduced to the Shepherd-Bremner construction as a special case. 
Similar to the Shepherd-Bremner construction, our construction also involves only one round of interaction. In the context of quantum cloud computing, the verifier (Alice) does not need to inform the prover of her purpose, and she can perform the verification protocol at any time, as long as she has access to the quantum server.
The difference is that, we may choose to construct the secret-encoded IQP circuit without depending on any error-correcting code. Therefore, our approach is intrinsically immune to the attack of Kahanamoku-Meyer's approach. 
Furthermore, this  model allows us to encode an unspecified number of secret strings, instead of relying only on a single secret string as in the Shepherd-Bremner protocol.
Intuitively, this significantly increases the difficulty for the prover to cheat, i.e., hacking multiple secret strings simultaneously with the same set of output bit strings. 
In practice, our protocol requires the quantum circuits to be in the supremacy regime, i.e. beyond the capability of classical simulation. Otherwise, the prover (Bob) can just use classical simulation to cheat, even though he is unable to find the secret strings.

In addition, we provide a general sampling method for an efficient estimation of any correlation function associated with a secret string, which gives the verifier a pre-determined value for checking. 
Moreover, in the special case where all angles are $\pi/8$ for local IQP gates, we discuss the possibility of ``quantizing" the values of the correlation functions through a connection with a family of Clifford circuits.
This gives an evaluation method for the correlation functions in this case based on the Gottesman-Knill algorithm~\cite{Gottesman98}. 
The efficient evaluation of correlation functions allows Alice not to rely on any error correcting codes.
However, we further show that for random IQP circuits, the majority of these correlation functions can be exponentially small, supported by numerical simulations together with an example on the random 2-local IQP circuits.
Consequently, we propose a heuristic strategy in which correlation functions with sizable values can be constructed by starting with a small system (Hamming weight), followed by a scrambling technique to expand the Hamming weight if necessary.

\section{Results and Methods}

\subsection{A general framework}

In Ref.~\cite{IQP08}, Shepherd and Bremner formulate an IQP-based verification protocol and give an explicit construction recipe for the IQP circuits used in the protocol. 
An IQP circuit of $n$ qubits can be represented by an $m$-by-$n$ binary matrix $\chi$, where each row represents a Pauli product. For example, a row vector $(1,1,0,0)$ represents $X_1 X_2$, where $X_i$ is the Pauli-$X$ acting on the $i$-th qubit. The Hamiltonian is the sum of $m$ Pauli operators, and the resulting IQP circuit is given by the time evolution of the Hamiltonian, i.e., $U_{\rm IQP} := e^{i\theta H}$. For example,
\begin{align}
    \chi = \begin{pmatrix}
    1 & 1 & 0 & 0 \\
    0 & 1 & 0 & 1
    \end{pmatrix} & \quad \Longrightarrow \quad  H = X_1 X_2 + X_2 X_4 \label{eq:matrix_to_hamiltonian}\\
    & \quad  \Longrightarrow \quad  U_{\rm IQP} = e^{i\theta X_1 X_2} e^{i\theta X_2 X_4} \ .
\end{align}
Here, one may choose $\theta = \pi/8$ as in the QRC construction~\cite{IQP08}. However, one may also consider the general cases where the angles of each term can be arbitrary, i.e., the angles may not necessarily be the same.

Our verification protocol is similar to the original Shepherd-Bremner construction, except that the encoding method is significantly extended. 
Below, following the physical picture previously introduced by the authors in Ref.~\cite{chen_experimental_2018}, we first present an overview of our protocol, and address several differences compared to the Shepherd-Bremner protocol.
An explicit construction recipe is given in Sec.~\ref{subsec:analysis}.
Let us consider a quantum circuit of $n$ qubits:
\begin{enumerate}

    \item[] \textbf{The quantum cloud verification protocol:}
    \item[\textbf{Step 1.}] Alice (the verifier) generates one, or multiple, $n$-bit random string(s) $\s:=(s_1,s_2, \cdots,s_n)^T \in\{0,1\}^n$ kept as a secret, where each string is associated with a Pauli product, $\Z_{\s} := Z^{s_1} \otimes \cdots \otimes Z^{s_n}$.
    

    \item[\textbf{Step 2.}] Based on the secret string(s), Alice designs a Hamiltonian $H$ consisting of a linear combination of Pauli-$X$ products. 
    
    
    
    
    \item[\textbf{Step 3.}] Alice then sends the classical description about the Hamiltonian $H$ to Bob (the prover) and asks him to apply the time evolution of $H$ to the state $\ket{0^n}$, where the angles of each term (i.e. the evolution time) are also determined by Alice. 
    
    \item[\textbf{Step 4.}] Bob should perform the quantum computation $U_{\rm IQP} |0^n\rangle$ accordingly and measure in the computational basis multiple times. After that, he returns the output bit strings to Alice.
    
    \item[\textbf{Step 5.}] Finally, Alice calculates the correlation function(s) $\ev{\Z_{\s}}:=\langle 0^n | U_{\rm IQP}^\dagger \Z_{\s} U_{\rm IQP} |0^n\rangle$ by classical means in Sec.~\ref{subsec:analysis}, and compares it with the results obtained from the bit strings given by Bob. 
\end{enumerate}





Note that in the Shepherd-Bremner protocol~\cite{IQP08}, Alice will instead check the probability bias $\P_{\s \perp}$, the probability of receiving bit strings that are orthogonal to $\s$, which is defined as,
\begin{align}
    \P_{\s \perp} := \sum_{\x \cdot \s = 0 } p(\x) \ ,
\end{align}
where $p(\x)$ is the output probability of the IQP circuit. However, as shown in Ref.~\cite{chen_experimental_2018}, the probability bias can be related to the correlation function as follows,
\begin{align}\label{eq:bias_and_cor_func}
    \P_{\s \perp} = \frac{1}{2} (\ev{\Z_{\s}} + 1) \ .
\end{align}
Therefore, these two measures of success are equivalent, but we choose to work with the correlation function, as it fits better our framework.

Furthermore, in the Shepherd-Bremner construction, only one secret string is considered for each time, and the Hamiltonian is constructed from a specific error-correcting code, the quadratic-residue code (QRC)~\cite{macwilliams1977book}, which can be regarded as a special instance in our framework. That is, if one constructs the Hamiltonian in Step~2 with QRC and chooses $\theta = \pi/8$, then our protocol reduces to the Shepherd-Bremner construction.

In order for the verification to work, Alice needs to know the value of the chosen correlation function in advance, which will be compared with the results from Bob's measurement data in Step~5. In the Shepherd-Bremner construction, $\ev{\Z_{\s}}$ is designed to always equal $1/\sqrt{2}$ (in terms of probability bias, 0.854) with only one specific $\s$, due to the properties of QRC.
In Sec.~\ref{subsec:analysis}, we present two methods for evaluating general correlation functions for IQP circuits, one corresponding to the most general case with arbitrary angles of each term, and the other corresponding to the case $\theta = \pi/8$. 
The former method is based on random sampling, while the latter is based on Gottesman-Knill algorithm~\cite{Gottesman98}.
With these two methods, Alice can calculate any $Z$-correlation functions as she wants, which allows her to test multiple secret strings.
We summarize the differences in Table.~\ref{tab:comparison}.

\begin{table}[t]
\begin{tabular}{|c|c|c|}
\hline
               & Shepherd-Bremner                                                          & our protocol         \\ \hline
circuit        & \begin{tabular}[c]{@{}c@{}}based on\\ quadratic residue code\end{tabular} & more general IQP circuits \\ \hline
secret string & single                                                                    & multiple             \\ \hline
$\theta$       & $\pi/8$                                                                   & arbitrary            \\ \hline
\end{tabular}
\caption{Comparison between our protocol and the Shepherd-Bremner protocol. }
\label{tab:comparison}
\end{table}



We now turn to discuss one possible way to incorporate multiple secret strings into the Hamiltonian in Step~2. 
Given secret strings $\s_1, \s_2, \cdots, \s_k$, Alice first generates the main part $H_M$ of the Hamiltonian, which has the property that every term in $H_M$ anti-commutes with $\Z_{\s_1}, \cdots\, \Z_{\s_k}$ simultaneously. This can be achieved in the following way. 
For a vector $\p \in \{0,1\}^n$, the associated Hamiltonian term is $\mathcal{X}_{\p} := X^{p_1} \otimes \cdots \otimes X^{p_n}$. Then $\mathcal{X}_{\p}$ anti-commutes with every $\Z_{\s_i}$ if $\p \cdot \s_i = 1$ for $i = 1, \cdots, k$. 
Therefore, the main part $H_M$ can be constructed from the solution space of this linear system.
To hide the secret strings, Alice would have to add a redundant part $H_R$, whose terms commute with every secret string $\s_i$. 
The redundant part can be similarly constructed from the solution space of the linear system $\p \cdot \s_i = 0$ for $i = 1, \cdots, k$. 
The whole Hamiltonian is $H = H_M + H_R$.
We remark that the values of the correlation functions depends only on the main part $H_M$ (see Sec.~\ref{subsec:analysis}). Therefore, Alice can add many redundant terms, to make the test harder.

Next, we make a further remark about the potential class of methods attacking our protocol, i.e., for Bob generating bit strings that can reproduce the value of the correlation function(s) without a quantum computer.
It would be possible if Alice's secret strings $\s$ are leaked to Bob; then, Bob could potentially evaluate the value of $\ev{\Z_{\s}}$, and output random bit strings according to the probability bias $\P_{\s \perp}$. 
Therefore, the security of the IQP-based protocols is based on the assumption that the secret string cannot be efficiently recovered from the Hamiltonian $H$. For the original Shepherd-Bremner protocol, such an attack has recently been found~\cite{kahanamoku-meyer_forging_2019} based on the properties of QRC.

Here our framework extends the encoding method in the protocol, making it immune to such kind of attack.
Furthermore, the number of secret strings is not revealed to Bob, instead of relying on a single secret string as in the previous protocol~\cite{IQP08}.
On the other hand, the IQP circuits constructed in our protocol is rather general. Although, a rigorous proof on the security of our protocol is missing, from a complexity-theoretic point of view, a general IQP circuit cannot be efficiently classically sampled, assuming some plausible conjectures~\cite{IQP10,bremner_average-case_2016}. 
In addition, the new features in our protocol should enhance the security compared to the original Shepherd-Bremner protocol.




\subsection{Detailed analysis}
\label{subsec:analysis}

\paragraph{Evaluating the correlation function given the secret string}
For Alice to evaluate correlation function(s), recall that relative to each string, the Hamiltonian $H=H_M+H_R$ can be divided into two parts: main part and redundant part. (i) The main part $H_M$ anti-commutes with $\Z_{\s}$, i.e., $\{ \Z_{\s}, H_M \} = 0$. (ii) The redundant part $H_R$ commutes with $\Z_{\s}$, i.e., $[H_R, \Z_{\s}] = 0$.

Due to these commuting properties relative to the secret strings, the value of the correction function only depends on the main part, i.e. (see Appendix~\ref{app:properties_correlation}), 
\begin{align}\label{eq:clifford}
    \ev{\Z_{\s}} = \mel{0^n}{e^{i 2\theta H_M} }{0^n} \ .
\end{align}
Note that one can arrive at a similar expression if we further relax the condition where the weight of each term can be uneven, e.g. $ H = \alpha X_1 X_2 + \beta X_2 X_4$.
On the other hand, the correlation function can be evaluated directly, if we confine the Hamming weight of the secret string to be sufficiently small, and the main part $H_M$ only acts on those qubits involved in the Pauli product $\Z_{\s}$.

\begin{figure}[t!]
    \centering
    \includegraphics[width = 0.5\textwidth]{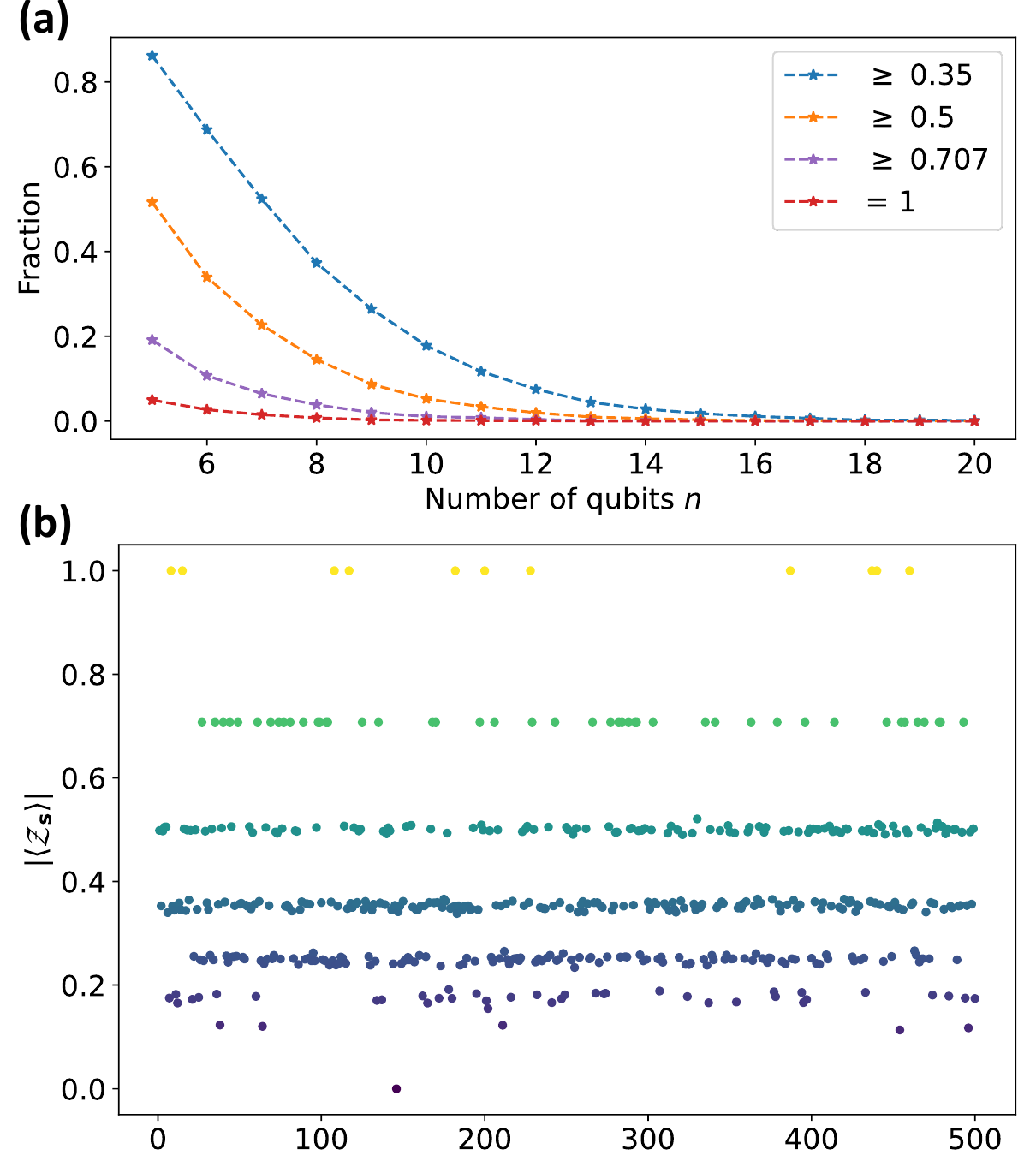}   
    \caption{Properties of the correlation functions. \textbf{(a)} Fractions of correlation functions of specific values versus the number of qubits $n$ in the IQP circuits. Data is obtained by searching over 10000 random IQP circuits for each $n$. \textbf{(b)} Correlation functions from 500 randomly generated 6-qubit IQP circuits and Pauli operators $\Z_{\s}$. The data fluctuation is because the correlation functions are obtained via random sampling.}
    \label{fig:data}
\end{figure}

For the general cases, the evaluation of the expression of $\ev{\Z_{\s}}$ can be achieved efficiently by sampling:
\begin{theorem}\label{thm:cor_func}
The correlation function of any Puali-Z product,  $\ev{\Z_{\s}}:=\langle 0^n | U_{\rm IQP}^\dagger \Z_{\s} U_{\rm IQP} |0^n\rangle$ associated with an IQP circuit, can be classically estimated to $\epsilon$ precision with probability $1 - \delta$, by taking  $\order{ \frac{1}{\epsilon^2} \log{\frac{2}{\delta}} }$ random samples.
\end{theorem}
To see why this theorem holds, one can apply local Hadamard gates to changing the main part $H_M$ to the $z$-basis, i.e., 
\begin{align}
    \ev{\Z_{\s}} = \frac{1}{2^n} \sum_{\x} \mel{\x}{(U_M^{(z)})^2}{\x} \ ,
\end{align}
where $U_M^{(z)}$ is the main part of the circuit with Pauli-$X$ replaced by Pauli-$Z$ (see Appendix~\ref{app:properties_correlation}). Each term in this summation can be efficiently calculated, with an absolute value bounded by 1. 
Then by the Chernoff bound argument, one can randomly sample bit strings $\x$, calculate $\mel{\x}{ (U_M^{(z)})^2 }{\x}$ for each bit string, and use the sample average to approximate $\ev{\Z_{\s}}$ to $\epsilon$ precision with probability $1 - \delta$, using $\order{ \frac{1}{\epsilon^2} \log{\frac{2}{\delta}} }$ samples. 
Thus, for a polynomially small correlation function, the precision $\epsilon$ can also be polynomially small, and this method is efficient with classical means. Note that using the same sampling method, Bob can also evaluate any correlation function efficiently even if the redundant part of the Hamiltonian is included. However, the correlation function(s) associated with the secret string(s) is hidden from Bob.

The above method is based on random sampling and approximation error will be incurred. 
In the special case of $\theta = \pi/8$, the correlation function can be evaluated exactly. 
Observe that $ \ev{\Z_{\s}} = \mel{0^n}{e^{i (\pi/4) H_M} }{0^n}$ actually corresponds to a transition amplitude of a Clifford circuit $e^{i (\pi/4) H_M}$. 
In this way, one can evaluate the correlation function efficiently using the Gottesman-Knill algorithm~\cite{Gottesman98}.
Specifically, the absolute value of $\ev{\Z_{\s}}$ is either $0$ or $2^{-g/2}$, where $0 \leq g \leq n$ is an integer determined by the stabilizer groups of $\ket{0^n}$ and $e^{i (\pi/4) H_M} \ket{0^n}$~\cite{aaronson_improved_2004}.
Note that in the Shepherd-Bremner construction, $g = 1$ for all QRC constructed IQP circuits.
We provide a example for the $g$-number in Appendix~\ref{app:properties_correlation}.
Also, see Fig.~\ref{fig:data}~(b) for an illustration of this `quantization' phenomenon with the simulation results of random 6-qubit instances; here, the randomness is from the random binary matrices $\chi$ for the IQP circuits and random secret strings.

\paragraph{Problem of random IQP circuits}
On the other hand, even though we have efficient classical algorithms for evaluating the correlation function to some additive error, it does not directly imply an effective solution to the problem. 
For a random instance of $H_M$, the value of the resulting correlation function could be small from the experimental point of view; this makes it difficult to distinguish from the uniform distribution. For example, for random 2-local IQP circuits of the form, $U = e^{i\frac{\pi}{8} \left(\sum_{i<j} w_{ij} X_i \otimes X_j + \sum_i v_i X_i \right) }$ with $w_{ij}, v_i \in \{0, 1, \cdots, 7\}$, we have the following theorem, bounding the occurring probability of large correlation functions:

\begin{theorem}\label{thm:random_2-local}
For the class of random 2-local IQP circuits, the probability of finding a polynomial-sized correlation function $\ev{\Z_{\s}}$ is exponentially small, i.e.,
\begin{align}
    \Pr_{U, \s} \left( \ev{\Z_{\s}}^2 \geq \frac{1}{k} \right) \leq \frac{3k}{2^n} \ ,
\end{align}
where $k = \poly(n)$ is a polynomial of $n$.
\end{theorem}
This is essentially due to the anti-concentration properties of IQP circuits~\cite{bremner_average-case_2016}, an important ingredient for proving the quantum computational supremacy of IQP sampling.
For the proof, we refer to Appendix~\ref{app:anti-concentration}.
We performed numerical simulation in Fig.~\ref{fig:data}~(a), which shows that the fraction of correlation functions that are larger than certain value decays quickly with the number of qubits. 
Theorem~\ref{thm:random_2-local} implies that for such kind of random Hamiltonians, it is not easy for Alice to keep secret strings associated with correlation functions with sizable values.

\begin{figure}[t!]
    \centering
    \includegraphics[width = 1\columnwidth]{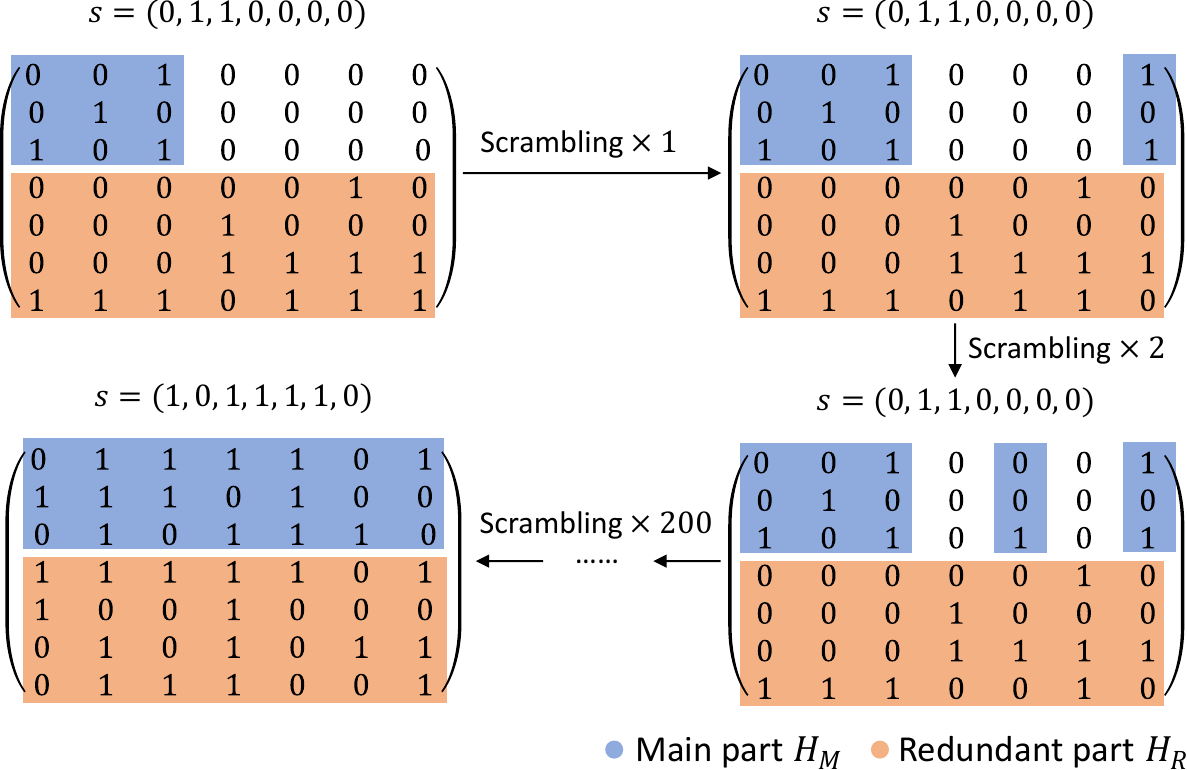}
    \caption{The scrambling process in the matrix representation. Here, $H_M$ (the blue block) initially acts only on first few qubits. After the first scrambling, the third column of the matrix is added to the last column, and the last entry of $\s$ is added to the third one, correspondingly. Similarly, after the second scrambling, the first column of the matrix is added to the fifth one, and the fifth entry of $\s$ is added to the first one. The resulting matrix after 200 times of scrambling is shown in the lower left corner, and the acting range of the main part extends to the whole circuit. }
    \label{fig:scrambling}
\end{figure}


\paragraph{Construction from an initially small main part}
Therefore, practically, Alice needs to carefully design the main part $H_M$ such that the correlation function is sufficiently away from zero.
In the case of $\theta = \pi/8$, it would be more desirable to have $g$ being zero or one, so that the correlation function becomes $1$ or $0.707$. 
However, as far as we are aware, there is no known efficient scheme of constructing an explicit Clifford circuit of the form ${e^{i (\pi/4) H_M} }$, for a fixed value of $g$. 
We anticipate that an systematic scheme can be designed to generate the desired main part with a given $g$ by leveraging the relation between IQP circuits and stabilizer formalism.
We leave it for future exploration.

Here, for the general case, we propose a heuristic method to construct the main part, which utilizes the feature that the correlation function with respect to any given secret string can be efficiently evaluated.
In our heuristic construction, Alice may start from secret strings of relatively small Hamming weight.
Then, a brute-force search is taken to find an $H_M$ that can yield an IQP circuit with sizable $\ev{\Z_{\s}}$ for every $\s$.
From Fig.~\ref{fig:data}~(a), the initial Hamming weight of the secret string should be limited to about 10, so that the search can stop in a reasonable time.
Then, one can extend the Hamming weight of the secret strings by the method of scrambling as previously introduced in Ref.~\cite{IQP08}. 
Specifically, the scrambling process refers to a column manipulation of the matrix $\chi$. 
For example, we can add the first column to the third in \eqref{eq:matrix_to_hamiltonian}, resulting in a new Hamiltonian $H = X_1 X_2 X_3 + X_2 X_4$.


Moreover, when all angles for the Hamiltonian terms are the same, the following theorem (as a consequence of Theorem~1 of \cite{IQP08}) implies that for IQP circuits, the scrambling process leaves the correlation function unchanged:
\begin{theorem}
\label{thm:scrambling_invariant}
Denote $\mathcal{C}_M$ as the linear subspace spanned by the column vectors of the main part, i.e., the matrix representation of $H_M$. When all $\theta$'s are identical, we can express the correlation functions in the following form:
\begin{align}\label{eq:scrambling_invariant}
    \ev{\Z_{\s}} = \frac{1}{2^{d}} \sum_{\c \in \mathcal{C}_{M}} \cos[2\theta (q - 2 |\c| )]  \ ,
\end{align}
where $d$ is the dimension of the linear subspace $\C_{M}$, $q$ is the number of terms in $H_M$ and $|\c|$ is the Hamming weight of $\c$. 
\end{theorem}
For completeness, we give a proof in Appendix~\ref{app:invariance_CF}.
From this theorem, one can see that the value of $\ev{\Z_{\s}}$ depends only on the linear subspace $\C_M$ since each term in the expression depends only on $\c \in \C_M$. 
Therefore, even if $H_M$ is scrambled, as long as the secret string is also scrambled accordingly, to preserve the inner product with rows in $\chi$, then the linear subspace $\mathcal{C}_M$ will remain unchanged and so is the value of $\ev{\Z_{\s}}$. 
We note that, this scrambling-invariance property is true even in the case where the angles are different; we discuss this point also in Appendix~\ref{app:invariance_CF}.

This scrambling-invariance property can be better explained with Fig.~\ref{fig:scrambling} as an example. 
In the upper left corner of Fig.~\ref{fig:scrambling}, $\s = (0,1,1,0,0,0,0)$ and the main part is the first 3 rows (i.e. $q = 3$). 
The linear subspace $\C_M$ is spanned by the column vectors in the blue block, namely $\vb{a}_1 = (0,0,1)^T$, $\vb{a}_2 = (0,1,0)^T$, and $\vb{a}_3 = (1,0,1)^T$ (i.e. $d = 3$). In this way, the vectors of the linear subspace takes the form $\c = \sum_{i = 1}^3 y_i \vb{a}_i$, with $y_i \in \{ 0,1 \}$. After the scrambling, one can check that each column in the new main part can be written as linear combination of the three vectors $\{ \vb{a}_1,\vb{a}_2,\vb{a}_3\}$ , which means the new linear subspace is the same as $\C_M$. 
So, the correlation function after the scrambling remains the same too. 
This scrambling process not only hides the secret, but also extends an initially small main part.
We remark that the scrambling invariance holds for the case of multiple secret strings as well, because those secrte strings all share with the same $\C_M$.

\section{Discussion}
Verification is important is various quantum computing models, such as blind quantum computation~\cite{Fitzsimons2017-verifiable-blind,xu_parallel_2020}, distributed quantum computation~\cite{buhrman2003distributed, Sheng_distributed_2017} and secure multi-party quantum computation~\cite{crepeau_2002_secure, Qiang_2017}.
In this paper, we have discussed a generalized model of IQP-based cryptographic verification protocol, where several features are introduced to achieve anti-forging of the quantum data. This model can be reduced to the Shepherd-Bremner protocol as a special case, but one can construct the secret without relying on the quadratic residue code, avoiding a class of attacks proposed by Kahanamoku-Meyer.

We remark that the applicability of our framework is not limited to IQP circuits. For example, one may design a quantum circuit of the following form: $U = U_R U_M$, where $U_R$ commutes with a certain observable $O$. 
In this way, we also have $\mel{0^n}{U^{\dagger} O U}{0^n} = \mel{0^n}{U_M^{\dagger} O U_M}{0^n}$ depends only on the main part. 
In the case that $U_M$ acts on a small number of qubits (i.e., a small main part), the expectation can be calculated by classical simulation.
Furthermore, one may consider an extension to quantum circuits like $\displaystyle U=U_{\rm shallow} V^{\dagger } V$, where $U_{\rm shallow}$ is a shallow sub-circuit and $V$ contains random unitary gates; of course, the overall gate order should be scrambled to the prover. In this case, the verifier can check any observable based on shallow part only. 

Returning to the IQP framework, the most imminent open question is a lack of a rigorous security proof of the general protocol. Practically, the protocol should be implemented in a scale beyond the capability of classical computing, i.e., in the regime of quantum advantage (supremacy). Furthermore, it is also necessary to take noises into account. These questions should be addressed with large-scale numerical simulations in future.

\textbf{Acknowledgement.---}
We thank Zhengfeng Ji and Michael Bremner for insightful discussions. This work is supported by the Natural Science Foundation of Guangdong Province (2017B030308003), the Key R\&D Program of Guangdong province (2018B030326001), the Science, Technology and Innovation Commission of Shenzhen Municipality (JCYJ20170412152620376 and JCYJ20170817105046702 and KYTDPT20181011104202253), National Natural Science Foundation of China (11875160 and U1801661), the Economy,Trade and Information Commission of Shenzhen Municipality (201901161512), Guangdong Provincial Key Laboratory(Grant No.2019B121203002).
BC thanks the support from the Sydney Quantum Academy, Sydney, NSW, Australia.





\bibliography{ref}

\newpage

\clearpage

\begin{appendix}

\section{Properties of the correlation functions}
\label{app:properties_correlation}

Let the main part and the redundant part of the IQP circuit be $U_M$ and $U_R$. Since each gate in the IQP circuit commutes with each other, we can assume without loss of generality that $U_{\rm IQP} = U_R U_M$, which gives $\ev{\Z_{\s}} = \mel{0^n}{U_M^{\dagger} U_R^{\dagger} \Z_{\s} U_R U_M}{0^n}$. From the fact that the redundant part $U_R$ commutes with $\Z_{\s}$, we have,
\begin{align}
    \ev{\Z_{\s}} = \mel{0^n}{U_M^{\dagger} \Z_{\s} U_M}{0^n} \ .
\end{align}
Furthermore, with the anti-commutation of $H_M$ and $\Z_{\s}$, we have $U_M^{\dagger} \Z_{\s} U_M = \Z_{\s} U_M^2$ and
\begin{align}\label{eq:corfunc_general}
    \ev{\Z_{\s}} = \mel{0^n}{U_M^2}{0^n} \ .
\end{align}
Then we apply Hadamard gates to change the basis,
\begin{align}
    \ev{\Z_{\s}} &= \frac{1}{2^n} \sum_{\x, \vb{y}} \mel{\x}{(U_M^{(z)})^2}{\vb{y}} \\
    &= \frac{1}{2^n} \sum_{\x} \mel{\x}{(U_M^{(z)})^2}{\x} \ , \label{eq:corfunc_diag}
\end{align}
where $U_M^{(z)}$ is the main part of the circuit with Pauli-$X$ replaced by Pauli-$Z$ and we have used the fact that $\mel{\x}{(U_M^{(z)})^2}{\vb{y}} = 0$ if $\x \neq \vb{y}$. Each term in the summation can be efficiently calculated by tracking the phase. By the Chernoff bound argument, $\ev{\Z_{\s}}$ can be approximated to $\epsilon$ precision with probability $1 - \delta$ using $\order{ \frac{1}{\epsilon^2} \log{\frac{2}{\delta}} }$ samples of $\x$, as stated in the main text.

Specifically, if $\theta = \pi/8$, then $U_M^2 = e^{i2\theta H_M}$, and \eqref{eq:corfunc_general} becomes,
\begin{align}
    \ev{\Z_{\s}} = \mel{0^n}{e^{i2\theta H_M}}{0^n} \ .
\end{align}
Here, $e^{i2\theta H_M}$ is a Clifford circuit since $2\theta = \pi/4$, and the correlation function can be exactly calculated by the Gottesman-Knill algorithm~\cite{Gottesman98} in this case.
Moreover, the absolute value of $\ev{\Z_{\s}}$ is either $0$ or $2^{-g/2}$, where $0 \leq g \leq n$ is the minimum number of different generators of the stabilizer groups of two states: $\ket{0^n}$ and $e^{i2\theta H_M} \ket{0^n}$. 
For example, if the generator associated to $\ket{0^n}$ is given by $\{ Z_1, Z_1 Z_2, Z_3, Z_3 Z_4 \}$, and that of $e^{i2\theta H_M} \ket{0^n}$ is given by $\{ Y_1 X_2, X_1 Y_2, Y_3 X_4, X_3 Y_4 \}$, then the number of different generators is 2; this can be seen by noting that the latter stabilizer group can be equivalently described by the following set of generators $\{ Y_1 X_2, Z_1 Z_2, Y_3 X_4, Z_3 Z_4 \}$.
The set of generators is not unique, and $g$ is the minimum over all possible generators associated to two states.

\section{The problem from anti-concentration}
\label{app:anti-concentration}

In Ref.~\cite{bremner_average-case_2016}, it is proved that for random 2-local IQP circuits of the form
\begin{align}\label{eq:IQP_form}
    U_{\rm IQP} = e^{i\frac{\pi}{8} \left(\sum_{i<j} w_{ij} X_i \otimes X_j + \sum_i v_i X_i \right) } \ ,
\end{align}
with $w_{ij}, v_i \in \{ 0, 1, \cdots, 7 \}$, the output probability is anti-concentrated. Specifically, the \emph{anti-concentration theorem} states that,
\begin{align}
    \bb{E}_{U} [p(\x)^2] \leq \frac{3}{2^{2n}}
\end{align}
for all $\x$, where $\bb{E}_{U}$ denotes a uniform average over all IQP circuits of the form of \eqref{eq:IQP_form}, that is over uniform choices of $w_{ij}$ and $v_i$. Then we have,
\begin{align}\label{eq:anti-concentration}
    \bb{E}_{U} \left[ \sum_{\x} p(\x)^2 \right] \leq \frac{3}{2^{n}} \ .
\end{align}

By definition, the correlation function can be written as,
\begin{align}\label{eq:corfunc_def}
    \ev{\Z_{\s}} = \sum_{\x} p(\x) (-1)^{\s \cdot \x} \ ,
\end{align}
which actually holds for a general quantum circuit. This means that $p(\x)$ is the Fourier transform of $\ev{\Z_{\s}}$, and that
\begin{align}
    p(\x) = \frac{1}{2^n} \sum_{\s} \ev{\Z_{\s}} (-1)^{\s \cdot \x} \ .
\end{align}
Then we can apply the Parseval's identity~\cite{Boolean},
\begin{align}
    \sum_{\x} p(\x)^2 = \frac{1}{2^n} \sum_{\s} \ev{\Z_{\s}}^2 \ ,
\end{align}
which gives,
\begin{align}
    \bb{E}_{U, \s} \left[ \ev{\Z_{\s}}^2 \right] &= \frac{1}{2^n} \sum_{\s} \bb{E}_U \left[ \ev{\Z_{\s}}^2 \right] \\
    &= \bb{E} \left[ \sum_{\x} p(\x)^2 \right] \\
    & \leq \frac{3}{2^n} \ .
\end{align}

The Markov's inequality gives the following bound, 
\begin{align}\label{eq:markov}
    \Pr_{U, \s}(\ev{\Z_{\s}}^2 \geq a) \leq \frac{\bb{E}_{U, \s} \left[ \ev{\Z_{\s}}^2 \right]}{a} \leq \frac{3}{a 2^n} \ ,
\end{align}
for $a > 0$. Setting $a = \order{1/\poly(n)}$, we have,
\begin{align}
\Pr_{U, s} \left( \ev{\Z_{\s}}^2 \geq \frac{1}{\poly(n)} \right) \leq \frac{3\poly(n)}{2^n} \ .
\end{align}
This means that for random 2-local IQP circuits, the probability that the correlation functions are polynomially small is exponentially small.

In practice, Alice will obtain the correlation function from Bob's data in the following way. Suppose Alice receives $T$ output strings from Bob, $\x_1, \cdots, \x_T$. For each string, Alice computes $(-1)^{\s \cdot \x_i}$, and the sample average
\begin{align}
    s_T := \frac{1}{T} \sum_i (-1)^{\s \cdot \x_i} 
\end{align}
gives an approximation of $\ev{\Z_{\s}}$. From the Chernoff bound argument, for $\epsilon$ precision and probability $1 - \delta$, $T = \order{ \frac{1}{\epsilon^2} \log{\frac{2}{\delta}} }$. In order for our protocol to be practical, $T$ should be a polynomial of $n$, which implies a polynomial precision $\epsilon = \order{1/\poly(n)}$. However, if $\ev{\Z_{\s}}$ itself is exponentially small, then Alice will not be able to distinguish the correct data from data obtained from uniform distribution, since polynomial precision in this case is not sufficient.

\section{Scambling invariance of correlation function}
\label{app:invariance_CF}
Now we want to prove that $\ev{\Z_{\s}}$ depends only on the linear subspace spanned by the matrix representation of $H_M$. First, denote $M$ as the matrix representation of $H_M$. \eqref{eq:corfunc_diag} gives,
\begin{align}
\ev{\mathcal{Z}_{\s}} &= \frac{1}{2^n} \sum_{\vb{y} \in \{ 0,1 \}^n } \mel{\vb{y}}{ \prod_{\p\in \row(M) } e^{i 2\theta_{\p} Z_{\p}} }{\vb{y}} \ ,
\end{align}
where $Z_{\p} := Z^{p_1} \otimes \cdots \otimes Z^{p_n}$. Then,
\begin{align}
\ev{\mathcal{Z}_{\s}} &= \frac{1}{2^n} \sum_{\vb{y} \in \{ 0,1 \}^n } \prod_{\p\in \row(M) } \exp(i 2\theta_{\p} (-1)^{\p\cdot \vb{y}} ) \\
&= \frac{1}{2^n} \sum_{\vb{y} \in \{ 0,1 \}^n } \exp(i \sum_{\p \in \row(M) } 2 \theta_{\p} (-1)^{\p\cdot \vb{y}} )  \ .
\end{align} 
Define $S$ as the scrambling matrix, which is product of elementary matrices in $\mathbb{F}_2$.
Then, after the scrambling, we have $\chi \to \chi \cdot S$, $M \to M \cdot S$ and $\s \to S^{-1} \s$, so that the inner-product relation between rows in $\chi$ and $\s$ is preserved.
Moreover, we have $\vb{p} \cdot \vb{y} = \vb{p} \cdot S S^{-1} \cdot \vb{y}$.
Then, denoting the new secret string as $\s' = S^{-1} \s$, the associated correlation function is given by,
\begin{align}
    \ev{\Z_{\s'}} &= \frac{1}{2^n} \sum_{\vb{y} \in \{ 0,1 \}^n } \exp(i \sum_{\p \in \row(M \cdot S) } 2 \theta_{\p} (-1)^{\vb{p} \cdot S S^{-1} \cdot \vb{y}} ) \\
    &= \frac{1}{2^n} \sum_{\vb{y} \in \{ 0,1 \}^n } \exp(i \sum_{\p \in \row(M) } 2 \theta_{\p} (-1)^{\p\cdot \vb{y}} ) \\
    &= \ev{\Z_{\s}} \ .
\end{align}
This proves the scrambling invariance of the correlation function.

If all angles are the same, then one can arrive at a similar theorem as Theorem~1 of Ref.~\cite{IQP08}.
Define $\vb{c}_{\vb{y}} := M \cdot \vb{y} $ to be an encoding of $\vb{y}$ under $M$. Recall that $M$ contains $q$ rows, so we can write
\begin{align}
    \c_{\vb{y}} = \begin{pmatrix} \p_1 \\ \vdots \\ \p_q \end{pmatrix} \cdot \y = \begin{pmatrix} \p_1 \cdot \y \\ \vdots \\ \p_q \cdot \y \end{pmatrix} \ ,
\end{align}
which means that each entry in $\c_{\y}$ equals $\p \cdot \y$ for $\p \in \row(M)$.
Then $\sum_{\p \in \row(M) } (-1)^{\p\cdot \vb{y}}$ equals the number of zeros in $\c_{\vb{y}}$ minus the number of ones (i.e., Hamming weight $|\c_{\vb{y}}| $), which gives,
\begin{align}
    \sum_{\p \in \row(M) } (-1)^{\p\cdot \vb{y}} = q - 2 |\c_{\vb{y}}| \ .
\end{align}

So, we arrive at,
\begin{align}
\ev{\Z_{\s}} &= \frac{1}{2^n} \sum_{\vb{y} \in \{ 0,1 \}^n } \cos[ 2\theta (q - 2 |\c_{\vb{y}}| ) ] \ ,
\end{align} 
where we used the fact that $\ev{\Z_{\s}}$ is real. Now, in the column picture, we can write $M = (\a_1, \cdots, \a_n)$, where $\a_i \in \col(M)$ is a column vector of length $q$. Thus, 
\begin{align}
    \c_{\y} = M \cdot \y = y_1 \a_1 + \cdots + y_n \a_n \ .
\end{align}
Suppose the dimension of $\C_{M}$ is $d$, and without loss of generality, assume $\{ \a_1, \cdots, \a_d \}$ forms a basis. Then,
\begin{align}
    \c = y_1 \a_1 + \cdots + y_{d} \a_{d}
\end{align}
for any $\c \in \C_{M}$. So the expression of $\ev{\Z_{\s}}$ becomes,
\begin{align}
    \ev{\Z_{\s}} &= \frac{1}{2^{n - d}} \sum_{y_{d+1}, \cdots, y_n} \left( \frac{1}{2^{d}} \sum_{\c \in \mathcal{C}_M} \cos[2\theta (q - 2 |\c| )] \right)
\end{align}
The first summation will give a factor of $2^{n - d}$, which cancels with $\frac{1}{2^{n -d}}$. So finally, it ends up giving,
\begin{align}
    \ev{\Z_{\s}} = \frac{1}{2^{d}} \sum_{\c \in \mathcal{C}_M} \cos[2\theta (q - 2 |\c| )]  \ .
\end{align}
Every term in the summation depends on the element $\c$, and therefore $\ev{Z_{\s}}$ depends only on the linear subspace $\mathcal{C}_{M}$.


\end{appendix}

\end{document}